\documentclass[10pt-default,11pt]{article}
\usepackage{amsmath}
\usepackage{amsfonts}
\usepackage{amssymb}
\usepackage{geometry}

\input{tcilatex}
\begin{document}

\bigskip

\bigskip

{\Large Probabilistic vs optical interpretation of quantum mechanics\medskip
\medskip }

\bigskip {\large \ Arkady L.Kholodenko}

\bigskip\ \textit{375 H.L.Hunter Laboratories, Clemson University,
Clemson,SC 29634-0973,USA\bigskip }

Abstract\bigskip

Although electrons and photons produce the same interference patterns \ in
the two-slit

experiments, the description of these patters is markedly different. This
difference was

analyzed by Bohm. Later on Sanz and Miret-Artes and others were able to
squeeze

the differences to zero. Fortunately, they left some room for developments
presented

in this Letter. We noticed that in the absence of sources the
electromagnetic field can

be represented by the complex scalar field. It is demonstrated that the same
fields are

being used in the non relativistic Schr\"{o}dinger equation. The connection
between the

electromagnetic and Schr\"{o}dinger fields allows to study the topology of
zero sets

(Chladni patterns) of \ Schr\"{o}dinger eigenfunctions. The existence of
these \ patterns

is contingent upon the existence of \ eigenvalues of multiplicity higher
than one. This is

permissible only in Schr\"{o}dinger's version of quantum mechanics. Presence
of

multiplicities is making quantum mechanical and topological entanglements
equivalent.

\bigskip

\textsl{Keywords:\bigskip \bigskip }

Maxwell's equations

Quantum field theory basics

Huygens' equivalence

Chladni patterns

Quantum entanglements

Knots and links \bigskip \bigskip

\textbf{1.Introduction\bigskip }

\bigskip\ 

If electrons and photons\footnote{%
And other massive particles} produce the same interference patterns in the
two- slit experiments [1-3], why then the optics formalism cannot be \
applied unchanged to electrons and heavier particles? What makes use of
Born's probabilistic \ interpretation of the wave function in quantum
mechanics superior to the intensity interpretation of the Maxwellian wave
function in optics? In particular, for the description of two-slit
experiments in which the results are visibly coinciding for light and
heavier particles? This issue was carefully investigated by David Bohm in
his classical monograph on quantum mechanics [4],pages 97-98. Reading of
these pages indicates that the differences exist just in a few places. The
updated theoretical comparison was recently made by Sanz and Miret-Artes, in
chapters 4 and 7 of [5]. From these chapters it follows that all objections
made by Bohm \ in the remaining few places can be removed. Fortunately, the
results of [5] along with those in references on which these results are
based, and even more recent ones, can still be improved. It is the purpose
of this Letter. As result, \ we are able to achieve the full correspondence
between the description of photons and of massive particles whose
Hamiltonians are manifestly time-independent. Since nowadays the
sophisticated quantum mechanical experiments are typically done optically
[6,7], the results of this letter may provide additional guidelines for
interpretation of these optical experiments and vice versa. \ In the light
of just mentioned correspondence, it makes sense to claim that our
understanding of subtleties of quantum mechanics is contingent upon our
understanding of optical formalism adopted for quantum mechanical needs. In
this Letter we shall discuss \ several features of optical formalism which
were not discussed yet in the context of quantum mechanics.

\ Even though the two-slits interference experiments produce the same
results in both optics and quantum mechanics, it is not immediately possible
to adopt word-for-word \ the optical formalism to quantum mechanics. This is
so because of the following. Maxwell's equations contain vector quantities,
like \textbf{E} and \textbf{H}. Besides, there is a polarization in optics
whose analog in quantum mechanics is spin [8]. The nonrelativistic Schr\"{o}%
dinger equation is spin-independent however. The two-slit interference
fringe pattern for monochromatic light depends strongly upon the light
polarization. For spinless particles in quantum mechanics used in two-slit
experiment this polarization effect is absent while in optics, depending on
polarization, there are four distinct cases to study. They were discovered
by Fresnel and Arago at the beginning of 19th century [9,10]. \ For the
record, these are: 1. Two rays of light polarized in the same plane. They
interfere like rays of ordinary (unpolarized) light. 2. Two rays polarized
at the right angles to each other. They do not interfere. 3. Two rays
originally polarized at the right angles and then brought into the same
plane of polarization. They do not produce interference pattern. 4 \ Two
rays originally polarized at the right angles, if derived from the same
linearly polarized wave and subsequently brought into the same plane, can
interfere.

These facts complicate the comparison \ between the two-slit interference\
experiments in optics and quantum mechanics. Fortunately, there are ways out
of these difficulties, e.g. those described in [5]. In this Letter we
describe still another methods. This task \ is accomplished in several
steps. The first step is discussed in Section 2. Its purpose is to introduce
some known results\ to be used in the reminder of this Letter.

\bigskip

\textbf{2. \ Derivation of the Schr\"{o}dinger equation for a single photon.}

\textbf{\bigskip }

We begin with writing down Maxwell's equations without sources and currents
in the vacuum. These are:

\begin{eqnarray}
\mathbf{\nabla }\cdot \mathbf{E} &=&0,\mathbf{\nabla }\times \mathbf{E}=-%
\frac{1}{c}\frac{\partial \mathbf{H}}{\partial t},  \notag \\
\mathbf{\nabla }\cdot \mathbf{H} &=&0,\mathbf{\nabla }\times \mathbf{H}=%
\frac{1}{c}\frac{\partial \mathbf{E}}{\partial t}.  \TCItag{1}
\end{eqnarray}%
In the above equations we keep only the speed of light $c$ while the rest of
constants we put equal to unity. They always can be restored \ if needed.
Incidentally, the situation described by (1) is in accord with that known in
standard quantum mechanics. This was emphasized by Bohm [4] and Bohm and
Hiley [11]. Waves in quantum mechanics do not have sources or sinks and,
accordingly, currents. In section 5 we provide arguments that the same can
be achieved for the Maxwellian fields if knotted/linked field configurations
are reinterpreted as charges. Next, we introduce the Riemann-Silberstein
complex vector 
\begin{equation}
\mathbf{F}_{\lambda }=\frac{1}{\sqrt{2}}(\mathbf{E}_{\lambda }+i\lambda 
\mathbf{H}_{\lambda }),  \tag{2}
\end{equation}%
where $\lambda =1(-1)$ corresponds to the positive (negative) helicity. The
concept of helicity is related to the concept of polarization \ which\ for
light is analogous to spin. All this is described in detail in [8].
Following Kobe [12], \ and Smith and Raymer [13,14],we temporarily suppress
the subscript $\lambda $ \ and introduce the energy density $\QTR{sl}{%
\varepsilon }$ as 
\begin{equation}
\mathbf{F}^{\ast }\cdot \mathbf{F}=\frac{1}{2}(\text{\textbf{E}}^{2}+\mathbf{%
H}^{2})\equiv \QTR{sl}{\varepsilon }.  \tag{3}
\end{equation}%
This result then allows us to introduce the photon wave function%
\begin{equation}
\Psi _{i}=\sqrt{\frac{1}{\mathcal{E}}}F_{i},  \tag{4}
\end{equation}%
Here i=1,2,3 labels Euclidean coordinates while $\mathcal{E}$ is defined in
(6). \ Using (4) we require%
\begin{equation}
\dsum\limits_{i=1}^{3}\dint d^{3}x\Psi _{i}^{\ast }\Psi _{i}=1  \tag{5}
\end{equation}%
with the total energy $\mathcal{E}$ defined by 
\begin{equation}
\dint d^{3}x\QTR{sl}{\varepsilon }=\mathcal{E}.  \tag{6}
\end{equation}%
By design, the wave function $\Psi _{i}$ is satisfying the Schr\"{o}dinger's
equation for the photon

\begin{eqnarray}
i\frac{\partial }{\partial t}\mathbf{F} &=&c\mathbf{\nabla }\times \mathbf{F,%
}\text{ }  \TCItag{7a} \\
\text{or, }i\hbar \frac{\partial }{\partial t}\mathbf{F} &=&ic\mathbf{%
p\times F,}\text{ }\mathbf{p=-i\hbar \nabla ,}  \TCItag{7b}
\end{eqnarray}%
provided that

\begin{equation}
\mathbf{\nabla }\cdot \mathbf{F}=0.  \tag{8}
\end{equation}%
Equations (7) and (8) are equivalent \ to Maxwell's equations (1) as
required. The continuity equation for the probability now reads%
\begin{equation}
\frac{\partial \rho }{\partial t}+\text{div}\cdot \mathbf{j=}0\mathbf{,}%
\text{ }\mathbf{j}=\frac{c}{\mathcal{E}}\mathbf{E}\times \mathbf{H=\frac{c}{%
\mathcal{E}}\mathbf{F}^{\ast }\times \mathbf{F},}\text{ }\rho \text{ }%
\mathbf{=}\dsum\limits_{i=1}^{3}\Psi _{i}^{\ast }\Psi _{i}.  \tag{9}
\end{equation}%
These results are perfectly fine as far as quantization of electromagnetic
field is of interest only. They are not exhibiting the universal connection
with quantum mechanics of particles though. \ They were designed for photons
only. This gives us an opportunity to describe such a connection.\bigskip
\bigskip \bigskip

\textbf{3. \ Electromagnetic field in the absence of currents and sources as}

\ \ \ \textbf{\ \ complex scalar field}\bigskip \bigskip

Following Green and Wolf [15,16], we notice that in a region of space free
of currents and charges the electromagnetic field is fully \ specified by
the single vector potential \textbf{A}. For such a case we can write: $%
\mathbf{E}=-\frac{1}{c}\frac{\partial }{\partial t}\mathbf{A\equiv -}\frac{1%
}{c}\mathbf{\dot{A},}$ $\mathbf{H=\nabla \times A.}$ For such cases the
vector potential \textbf{A} can be rigorously derived from a single (in
generally complex) scalar field potential $V(\mathbf{x},t)$. It can be shown
that the total energy $\mathcal{E}$, defined in (3) and (6), can be
rewritten in terms of $V(\mathbf{x},t)$ as 
\begin{equation}
\dint d^{3}x\QTR{sl}{\varepsilon }\equiv \frac{1}{2}\dint d^{3}x\text{ }(%
\frac{1}{c^{2}}\dot{V}\dot{V}^{\ast }+\mathbf{\nabla }V\cdot \mathbf{\nabla }%
V^{\ast })=\mathcal{E}  \tag{10}
\end{equation}%
Accordingly, the Poynting flux $\ \mathbf{j}$\textbf{\ \ }of \
electromagnetic energy density\textbf{\ }is given by 
\begin{equation}
\mathbf{j}=-\frac{1}{2}(\dot{V}^{\ast }\mathbf{\nabla }V+\dot{V}\mathbf{%
\nabla }V^{\ast }).  \tag{11}
\end{equation}%
Therefore, the\ analog of the continuity equation (9) now reads as 
\begin{equation}
\frac{\partial \QTR{sl}{e}}{\partial t}+\text{div}\cdot \mathbf{j=}0. 
\tag{12}
\end{equation}%
It should be clear that the obtained results are equivalent to \ those which
are presented in section 2. The advantage of having these results rewritten
with help of scalar potential lies in the opportunity of bring them into
correspondence with quantum mechanics. In fact, as we soon demonstrate, such
scalar form of Maxwell's equations allows us to accomplish much more. To
begin, we represent $V(\mathbf{x},t)$ in the form

\begin{equation}
V(\mathbf{x},t)=\dint d^{3}k[\alpha (\mathbf{k},t)\cos (\mathbf{k}\cdot 
\mathbf{x})+\beta (\mathbf{k},t)\sin (\mathbf{k}\cdot \mathbf{x})]\equiv
\dint d^{3}kV(\mathbf{k},t),  \tag{13}
\end{equation}%
while at the same time, 
\begin{equation}
\mathbf{A}(\mathbf{x},t)=\dint d^{3}k[\mathbf{a}(\mathbf{k},t)\cos (\mathbf{k%
}\cdot \mathbf{x})+\mathbf{b}(\mathbf{k},t)\sin (\mathbf{k}\cdot \mathbf{x}%
)]=\dint d^{3}k\mathbf{A}(\mathbf{k},t).  \tag{14}
\end{equation}%
From electrodynamics it is known that the vector potential \textbf{A} is
satisfying the vector wave equation 
\begin{equation}
\frac{1}{c^{2}}\frac{\partial ^{2}}{\partial ^{2}t}\mathbf{A}-\nabla ^{2}%
\mathbf{A}=0.  \tag{15a}
\end{equation}%
Using (14) in (15a) leads to%
\begin{equation}
\frac{1}{c^{2}}\frac{\partial ^{2}}{\partial ^{2}t}\mathbf{A}(\mathbf{k},t)+%
\mathbf{k}^{2}\mathbf{A}(\mathbf{k},t)=0.  \tag{15b}
\end{equation}%
By comparing \ equations (13),(14), it follows that 
\begin{equation}
\frac{1}{c^{2}}\frac{\partial ^{2}}{\partial ^{2}t}V(\mathbf{k},t)+\mathbf{k}%
^{2}V(\mathbf{k},t)=0  \tag{16a}
\end{equation}%
and, therefore, 
\begin{equation}
\frac{1}{c^{2}}\frac{\partial ^{2}}{\partial ^{2}t}V(\mathbf{x},t)-\nabla
^{2}V(\mathbf{x},t)=0.  \tag{16b}
\end{equation}%
Next, suppose that the solution of (16b) can be represented in the form%
\begin{equation}
V(\mathbf{x},t)=\mathfrak{v}(\mathbf{x})\exp \{i\phi (\mathbf{x},t)\},\text{ 
}\phi (\mathbf{x},t)=k\Phi (\mathbf{x})-\omega t.  \tag{17}
\end{equation}%
It should be noted that such a representation is physically motivated \ by
its direct connection with Huygens' principle. Details are best explained in
monographs by Hilbert and Courant [17] \ and by Luneburg [18]. \ Some (but
not all!) of their ideas were subsequently developed by Maslov [19].
Different/alternative interpretation of (17) adopted for quantum mechanics
was independently developed by Bohm in his Bohmian mechanics [11]. Using
(17) in (16b) the following two equations are obtained: 
\begin{equation}
\left( \nabla \Phi \right) ^{2}-\frac{1}{k^{2}\mathfrak{v}}\nabla ^{2}%
\mathfrak{v}=n^{2},  \tag{18a}
\end{equation}%
\begin{equation}
\nabla \Phi \cdot \nabla \mathfrak{v+}\frac{1}{2}\left( \nabla ^{2}\Phi
\right) \mathfrak{v}=0.  \tag{18b}
\end{equation}%
Here $k=c/\omega ,$ $n^{2}=1.$ In the case of quantum mechanics, typically, $%
n^{2}=n^{2}(\mathbf{x}).$This fact \ will be discussed further below. For
now, however, consider both of these equations in the limit $k^{2}$ $%
\rightarrow \infty .$ Such a limit is typical for the phenomena described by
methods of geometrical optics [20]. In this limit (18a) is converted into : 
\begin{equation}
\left( \nabla \Phi \right) ^{2}=n^{2}\text{ .}  \tag{19a}
\end{equation}%
Physics behind equations (18a) and (18b) is predetermined by the geometric
optics limit. Specifically, in this limit surfaces $\Phi =const$ represent
the \textsl{wave fronts} while their duals represent \textsl{\ light rays. }%
These are orthogonal to \ the wavefronts. Following [19,20], we notice that:
a) (19a) is known as the \textsl{eikonal equation}, b)(19b) is known as 
\textsl{transport equation}. This equation can be simplified with help of
the following arguments. If $\frac{\partial }{\partial \tau }$ denotes the
differentiation along a particular ray, then according to [15,19,20] we
write, 
\begin{equation}
\frac{\partial }{\partial \tau }\cdot \cdot \cdot =\nabla \Phi \cdot \nabla
\cdot \cdot \cdot .  \tag{20}
\end{equation}%
Displayed identity allows us to rewrite (18b) as 
\begin{equation}
\frac{\partial \mathfrak{v}}{\partial \tau }+\frac{1}{2}\left( \mathfrak{v}%
\nabla ^{2}\Phi \right) =0.  \tag{19b}
\end{equation}%
Integration of the last equation is straightforward and is yielding the
result:%
\begin{equation}
\mathfrak{v}(\tau )=\mathfrak{v}(\tau _{0})\exp [-\frac{1}{2}%
\dint\limits_{\tau _{0}}^{\tau }d\tau ^{\prime }\nabla ^{2}\Phi ].  \tag{19c}
\end{equation}%
Next, following [20], chr.7, we notice that the ray trajectory $\mathbf{x}$($%
\tau )$ can be derived from the equations of motion 
\begin{equation}
\frac{d\mathbf{x}}{d\tau }=\mathbf{\nabla }\Phi .  \tag{21}
\end{equation}%
By combining equations(19b) and (21) the solution, equation (19c), can be
rewritten as 
\begin{equation}
\mathfrak{v}(\mathbf{x}(\tau ))=\mathfrak{v}(\mathbf{x}(0))\exp [-\frac{1}{2}%
\dint\limits_{\tau _{0}}^{\tau }d\tau ^{\prime }\nabla ^{2}\Phi (\mathbf{x}%
(\tau ^{\prime }))].  \tag{22}
\end{equation}%
Notice next that$\mathbf{x}(\tau )=\mathbf{x}(\mathbf{x}_{0},\tau )$ so that 
$\mathbf{x}(\mathbf{x}_{0},\tau _{0})=\mathbf{x}_{0}.$ This observation
allows us to introduce the Jacobian $\ J$ as follows 
\begin{equation}
J=\det (\frac{\partial x^{i}(\mathbf{x}_{0},\tau )}{\partial x_{0}^{j}}). 
\tag{23}
\end{equation}%
In view of (21), we can obtain as well%
\begin{equation*}
\frac{1}{J}\frac{dJ}{d\tau }=\nabla \cdot \nabla \Phi =\nabla ^{2}\Phi (%
\mathbf{x}(\tau ))
\end{equation*}%
implying%
\begin{equation}
J(\mathbf{x}_{0},\tau )=\exp [\dint\limits_{\tau _{0}}^{\tau }d\tau ^{\prime
}\nabla ^{2}\Phi (\mathbf{x}(\tau ^{\prime }))]\equiv J(\mathbf{x}(\tau )). 
\tag{24}
\end{equation}%
By combining (22) and (24) we finally obtain: 
\begin{equation}
\mathfrak{v}(\mathbf{x}(\tau ))=\frac{\mathfrak{v}(\mathbf{x}(\tau _{0}))}{%
\sqrt{J(\mathbf{x}(\tau ))}}.  \tag{25}
\end{equation}%
In view of this result, it is always possible to normalize $\mathfrak{v}(%
\mathbf{x}(\tau )),$that is to require%
\begin{equation}
\dint_{\Delta }d^{3}x\mathfrak{v}^{2}=1  \tag{26}
\end{equation}%
where $\Delta $ is the domain of integration determined by a particular
problem to be solved. According to [15], the energy density $\QTR{sl}{%
\varepsilon }$ \ defined in (10), can be represented in view of (19a) as \ 
\begin{equation}
\varepsilon =\frac{1}{2}[k^{2}\mathfrak{v}^{2}+\frac{k^{2}}{n^{2}}\mathfrak{v%
}^{2}(\left( \nabla \Phi \right) ^{2}+\frac{1}{k^{2}}(\nabla \ln \mathfrak{v}%
)^{2})]\rightarrow _{k\rightarrow \infty }k^{2}\mathfrak{v}^{2}.  \tag{27}
\end{equation}%
In the limit $k\rightarrow \infty $ the analog of the \ wavefunction density 
$\rho $ , (9), is given now by 
\begin{equation}
\rho =\frac{\varepsilon }{\mathcal{E}}\rightarrow _{_{k\rightarrow \infty }}%
\frac{\mathfrak{v}^{2}}{\dint d^{3}x\mathfrak{v}^{2}}\rightarrow \Psi ^{\ast
}\Psi ,\text{ }\Psi =\mathfrak{v}(\mathbf{x}(\tau ))\exp \{i\phi (\mathbf{x}%
,t)\}.  \tag{28}
\end{equation}%
Substitution of the ansatz (17) into expression for the energy current (11)
results in 
\begin{equation}
\frac{\mathbf{j}}{\mathcal{E}}=\frac{1}{2}\mathfrak{v}^{2}(\mathbf{x})%
\mathbf{\nabla }\Phi .  \tag{29}
\end{equation}%
The associated Hamilton-Jacobi (H-J) equation is given by (19a). In the
limit $k\rightarrow \infty $ its solutions are those for the Schr\"{o}dinger
equation. This fact is well known from the WKB theory where $k\rightarrow
\infty $ limit\ is the same as $\hbar \rightarrow 0$ limit (the classical
limit). We shall recover below Schrodinger's equation without recourse to
the $k\rightarrow \infty $ (or $\hbar \rightarrow 0)$ limit.

\bigskip .

\bigskip

\textbf{4. \ The role of electromagnetic vs vector \ and complex scalar
fields\bigskip\ in }

\ \ \ \ \ \textbf{Schr\"{o}dinger's quantum mechanics}

\bigskip

It is well known that \ the source-free Maxwell's field \ equations can be
recast in terms of the equations for the massless Dirac fields\footnote{%
With Dirac matrices being replaced \ by matrices for spin 1 fields} [8,21],
in terms of the massless 4-component vector fields [22], and, \ in view of
the results of this Letter, in terms of the massless complex scalar fields.
There is still \ \ another option. A consistent relativistic quantum
mechanics of spin 0 and 1 bosons can be developed with help of the \
Duffin-Kemmer equation [22-24]. This is Dirac-like 1st order equation in
which the Dirac matrices are replaced by $\beta $ matrices obeying
commutation relation similar to those for the Dirac matrices. In dealing
with the (anti) self -dual electromagnetic fields the twistorial
interpretation of equations for these fields is also possible [25]. The
comparison between Maxwell's fields, the complex scalar, the Dirac and the
vector fields is seemingly possible only for massless versions of these
fields. The twistorial formalism apparently also excludes uses of the
massive fields. Inclusion of masses into results mentioned thus far does not
cause additional difficulties though. \ All listed massive fields are
reducible to the massless \ ones as will be explained below, in section 5.

To our knowledge, the interpretation of electromagnetic fields in terms of
the complex scalar fields was not in use in physics literature thus far. \
The original paper by Roman [26] written in 1959 apparently was left
unnoticed. This could be seen from the series of papers culminating in
[23]-all using much more cumbersome Duffin-Kemmer formalism. Without
repeating their content, and to avoid overlaps, we \ still would like to add
some comments now. In standard field-theoretic notations \ [8,22] (making
for a moment all constants equal to unity) the Lagrangian $\mathcal{L}$ for
the complex massless scalar fields is given by, e.g. see [22], page 32, 
\begin{equation}
\mathcal{L}[\varphi ,\varphi ^{\ast }]=\dsum\limits_{n}\frac{\partial
\varphi }{\partial x^{n}}\frac{\partial \varphi ^{\ast }}{\partial x_{n}}. 
\tag{30a}
\end{equation}%
By varying the fields $\varphi $ and $\varphi ^{\ast }$ in the action $%
\mathcal{A}$ defined by 
\begin{equation}
\mathcal{A}=\dint d^{4}x\mathcal{L}[\varphi ,\varphi ^{\ast }]  \tag{30b}
\end{equation}%
while assuming that these fields are independent and \ nicely decaying at
infinity leads to the following equations of motion%
\begin{eqnarray}
\square \varphi &=&0,  \TCItag{31a} \\
\square \varphi ^{\ast } &=&0.  \TCItag{31b}
\end{eqnarray}%
Here the d'Alembertian $\square $ is defined as usual: $\square =\frac{%
\partial ^{2}}{\partial ^{2}t}-\nabla ^{2},$ with $\nabla ^{2}$ being the
3-dimensional Laplacian$.$ In his work [26] Roman was not interested in the
standard field-theoretic analysis of $\mathcal{L}[\varphi ,\varphi ^{\ast }]$%
, e.g. that done in [22] on page 32. He was interested in proving that the
gauge transformations of the Maxwellian fields \ rewritten in the formalism
of complex scalar fields will keep the action $\mathcal{A}$ form-invariant
(up to the total divergences vanishing at the boundary of space-time). The
purpose of Roman's work was in proving this result. With this result proven,
the task of this Letter is different.

Given that the \ action $\mathcal{A}$ is \ gauge-invariant, we apply the
standard field-theoretic treatment to (30) with purposes which will become
obvious upon reading. From [22] we find the time component $T^{00}$ of the
energy-momentum tensor (that is the energy density). It is given by 
\begin{equation}
T^{00}=\frac{\partial \varphi }{\partial t}\frac{\partial \varphi ^{\ast }}{%
\partial t}+\mathbf{\nabla }\varphi \cdot \mathbf{\nabla }\varphi ^{\ast }. 
\tag{32}
\end{equation}%
$T^{00}$ coincides with $\QTR{sl}{\varepsilon }$ defined by (10) where we
temporarily put $c=1$. Accordingly, the momentum density $T^{0\alpha }$
given by 
\begin{equation}
T^{0i}=-(\frac{\partial \varphi ^{\ast }}{\partial t}\nabla _{i}\varphi +%
\frac{\partial \varphi }{\partial t}\nabla _{i}\varphi ^{\ast }),i=1,2,3, 
\tag{33}
\end{equation}%
up to a constant coincides with the flux $\ \mathbf{j}$ defined in (11).
Evidently, the continuity equation (12) is nothing but the law of
conservation of the energy-momentum tensor [8]%
\begin{equation}
\frac{\partial }{\partial x^{\mu }}T_{\nu }^{\mu }=0.  \tag{34}
\end{equation}%
For the record, we use the Minkowski- space metric tensor $g_{\mu \nu }$
with signature $(+,-,-,-),$ so that $g_{\mu \nu }=g_{{}}^{\mu \nu }$ . For
the 4-vector $a_{\mu }$ we have $a_{\mu }=g_{\mu \nu }a^{\nu },$ etc.

Because we are dealing with the complex scalar field, there is also a
current vector $J_{\mu }$ responsible for carrying the charge (recall that
the description of charges and currents associated with them is always
associated with the existence of global gauge symmetry linked with use of
complex scalar fields\ instead of real ones [22]). Mathematically, the
conservation of current $J_{\mu }$ is expressible by analogy with (35) as%
\begin{equation}
\frac{\partial }{\partial x^{\mu }}J^{\mu }=0.  \tag{35}
\end{equation}%
Explicitly, the current $J_{\mu }$ is given by 
\begin{equation}
J_{\mu }=i(\varphi ^{\ast }\frac{\partial \varphi }{\partial x^{\mu }}-\frac{%
\partial \varphi ^{\ast }}{\partial x^{\mu }}\varphi ).  \tag{36}
\end{equation}%
Let $Q=\dint d^{3}xJ_{0}$ be the total charge. Then (35) is the continuity
equation associated with the charge conservation. It is well known [8] that $%
J_{0}$ is not always positively defined quantity. This fact \ is caused by
the observation that at any given time $t$\ both $\varphi $ and $\frac{%
\partial \varphi }{\partial x^{0}}$ may independently have arbitrary values.
\ This fact is important because of the following. \ For the sake of
generality and comparison, we include the mass $m^{2}$ term into both
equations (31a,b) \ thus converting them into the Klein-Gordon (K-G)
equations [22], page 32, 
\begin{eqnarray}
\left( \square +m^{2}\right) \varphi &=&0,  \TCItag{37a} \\
\left( \square +m^{2}\right) \varphi ^{\ast } &=&0.  \TCItag{37b}
\end{eqnarray}%
Since times of Schr\"{o}dinger's \ discovery \ of the equation bearing his
name, the K-G equation was considered as relativistic analog of the Schr\"{o}%
dinger's equation. In analogy with non relativistic case by multiplying
(37a) by $\varphi $ and (37b) by $\varphi ^{\ast }$ and subtracting (37b)
from (37a) we are repeating the same steps as for the non relativistic Schr%
\"{o}dinger equation in order to obtain the continuity equation (35). \ In
the nonrelativistic case (that is for the Schr\"{o}dinger equation) this
procedure yields: $j_{0}=\rho =\psi ^{\ast }\psi $ and $\vec{j}=\frac{\hbar 
}{2mi}(\psi ^{\ast }\mathbf{\nabla }\psi -\psi \mathbf{\nabla }\psi ^{\ast
}).$ The continuity equation in the nonrelativistic case is in its standard
form: $\frac{\partial }{\partial t}\rho +\mathbf{\nabla }\cdot \vec{j}=0.$
To compare this result with equations (35), (36) we must multiply (35) by $%
\frac{\hbar }{2m}.$ Then, we obtain respectively $J_{0}=\frac{i\hbar }{2m}%
(\varphi ^{\ast }\frac{\partial \varphi }{\partial x^{0}}-\frac{\partial
\varphi ^{\ast }}{\partial x^{0}}\varphi )$ and $J_{k}=\frac{\hbar i}{2m}%
(\varphi ^{\ast }\frac{\partial \varphi }{\partial x^{k}}-\frac{\partial
\varphi ^{\ast }}{\partial x^{k}}\varphi )=-$ $j_{k}$ , $k=1,2,3$. Thus,
even though (up to a sign) the currents $\vec{J}$ and $\vec{j}$ \ coincide,
the densities $j_{0}$ and $J_{0}$ are noticeably different. This fact
matters when the time-dependent problems are discussed for the K-G fields.
Because of the nonpositivity of $J_{0}$ the full time-dependent K-G equation
was discarded \ as the relativistic analog of the Schr\"{o}dinger equation.
In the time-independent case the situation is not so dramatic. Specifically,
suppose that the field $\varphi $ in (37a) can be written as $\varphi (%
\mathbf{x},t)=\psi (\mathbf{x})\exp (-i\omega t),$ then, the K-G equation is
converted into the Helmholtz equation 
\begin{equation}
\left( \mathbf{\nabla }^{2}+\tilde{m}^{2}\right) \psi =0.  \tag{38}
\end{equation}%
Accordingly, the field $\varphi ^{\ast }$ can be written now as $\varphi
^{\ast }(\mathbf{x},t)=\psi (\mathbf{x})\exp (i\omega t)$. In view of these
results, $J_{0}$ now acquires the form: $J_{0}=\frac{\hbar \omega }{m}(\psi
^{\ast }\psi ).$ At the same time\ $J_{k}=0.$ This result allows us to study
all kinds of stationary K-G equations [$27$] all making physical sense. The
obtained result raises the following question. In section 3 we demonstrated
that the continuity equation (12) is exactly the same as the energy-momentum
conservation equation (34). At the same time, the continuity equation (9)
for the photon is the same as the continuity equation (12) for the complex
scalar field. This means that this equation can be used instead of the
continuity equation (35) \ for development of the time-dependent quantum
mechanical formalism for the complex scalar K-G field. It is appropriate at
this point to mention that historically the massive complex scalar field was
used in Yukawa theory of strong interactions where it is known as pi meson
field. By extending results by Harish-Chandra [23] for this field written in
the language of Duffin-Kemmer formalism already mentioned, Tokouoka [28] had
studied in detail the meson-nucleon interactions. Much more recent results \
are discussed, for example, in [29] and references therein. Since the meson
-nucleon interactions are described nowadays with help of quantum
chromodynamics (QCD), the complex scalar field acts effectively as the
abelianized version of the non Abelian Yang-Mills gauge field. More on this
is discussed in the next section where we shall also explain how to get rid
of the mass term in the K-G equation. According to [27], page 99, every
spinor component of the Dirac equation with nonzero mass is satisfying the
massive K-G equation. \ In the zero mass limit this fact creates the
equivalence class between the massless K-G and Dirac equations. Apparently
components of equations for higher spin particles, e.g. spin-2 gravitons,
also belong to the same equivalence class [8,22,27]. In this section we
still need to demonstrate how results of sections 3 and 4 are connected with
the non relativistic time-independent Schr\"{o}dinger equation (TISE). In
the next section we shall present evidence that such an equation also
belongs to the same equivalence class as the rest of basic equations for
integer and half integer spins.

To connect results of sections 3 and 4 with TISE, just described treatment
of the stationary K-G equation is the most helpful. After the mass term is
eliminated in this equation\footnote{%
As described below}, the K-G equation is converted into the wave equation
(31). This fact allows us to use the results of Schr\"{o}dinger's second
foundational paper on quantum mechanics [30], pages 13-40. In (31) we
replace the speed of light $c$ by $u=c/n,$ where $n=n(x,y,z)$ is the
effective refractive index. Next, we use \ the same ansatz $\varphi (\mathbf{%
x},t)=\psi (\mathbf{x})\exp (-i\omega t)$ for the wave function. Next we
replace the dispersion relation $\omega =ck$ for electromagnetic waves in
the vacuum by $\omega =uk$ and use the de Broglie-type relation $k=p/\hbar $
along with the fact that for the mechanical system $E=\mathbf{p}^{2}/2m$ +$V$
. Thus we obtain: $\mathbf{p}^{2}$ =2m$(E-V).$ With the help of these
results the TISE follows: 
\begin{equation}
\lbrack -\frac{\hbar ^{2}}{2m}\nabla ^{2}+V-E]\psi =0.  \tag{39}
\end{equation}%
It follows immediately from the wave equation \ (31) under conditions just
described. The normalized result for $J_{0}$ can be used immediately so that
the continuity \ equation (35) works in this case. Nevertheless, the
question remains: what to do with the continuity equation (34)? This
equation is used for development of quantum mechanics of photons. Can it be
used \ in the nonrelativistic case of stationary Schr\"{o}dinger equation?
Equations (31)-(34) suggest that this could be possible. Nothing meaningful
is obtained straightforwardly though. Indeed, by using the K-G ansatz for
the wave function in (32) and (33) and restoring $c$ in these equations
yields:%
\begin{eqnarray}
T^{00} &=&\left( \frac{\omega }{c}\right) ^{2}\psi ^{\ast }\psi +\mathbf{%
\nabla }\psi \cdot \mathbf{\nabla }\psi ^{\ast },  \TCItag{40a} \\
T^{0i} &=&0.  \TCItag{40b}
\end{eqnarray}%
The result obtained for $T^{00}$ is different from that for $J_{0}.$ Unlike
[16], we are not going to look for arguments in favor of $\mathbf{\nabla }%
\psi \cdot \mathbf{\nabla }\psi ^{\ast }\simeq \left( \frac{\omega }{c}%
\right) ^{2}$ relation$.$This was done already in section 3 where results
were obtained in the regime of geometrical optics. Instead, now we are going
to use Schr\"{o}dinger's ideas again. This time we are going to use methods
developed in his 1st paper on quantum mechanics [30], pages 1-12, as well as
from the already discussed 2nd paper. In particular, we again replace $c$ by 
$u$ in (40a) and then replace $\left( \frac{\omega }{u}\right) ^{2}$ by $%
\left( \frac{p}{\hbar }\right) ^{2}.$\ \ Next, by looking at (40a) we require%
\begin{equation*}
\mathbf{\nabla }\psi \cdot \mathbf{\nabla }\psi ^{\ast }=\left( \frac{p}{%
\hbar }\right) ^{2}\psi ^{\ast }\psi
\end{equation*}%
and use $\ \left( \frac{p}{\hbar }\right) ^{2}=\frac{2m(E-V)}{^{\hbar ^{2}}}%
. $ \ These results can be rewritten in terms of the H-J equation if we
temporarily use only real valued functions: $\psi (\mathbf{x})=\psi ^{\ast }(%
\mathbf{x})$%
\begin{equation}
\left( \nabla \psi \right) ^{2}=\frac{2m(E-V)}{^{\hbar ^{2}}}\psi ^{2} 
\tag{41}
\end{equation}%
Notice that this H-J equation contains explicitly Planck's constant $\hbar $
while the H-J equation used in the semiclassical WKB calculations is by
design $\hbar -$ independent. Nevertheless, (41) coincides exactly with the
equation ($1^{\prime \prime })$ of Schr\"{o}dinger's 1st paper on quantum
mechanics, [30], pages 1-12. \ This difference has a profound effect on the
rest of Schrodinger's calculations. It allows\ him and us to restore \ the
stationary Schrodinger equation without any approximations. For this
purpose, Schr\"{o}dinger introduces the functional $J[\psi ]$ 
\begin{equation}
J[\psi ]=\frac{1}{2}\dint d^{3}x[\left( \nabla \psi \right) ^{2}-\frac{%
2m(E-V)}{^{\hbar ^{2}}}\psi ^{2}]  \tag{42a}
\end{equation}%
which he is minimizing under the subsidiary condition%
\begin{equation}
\dint d^{3}x\psi ^{2}=1.  \tag{42b}
\end{equation}%
Minimization produces the stationary Schr\"{o}dinger equation (39) as
anticipated. Thus, the continuity equations (34) and (35) \ both can be used
in conjunction with the nonrelativistic stationary Schr\"{o}dinger's
equation (39).

\bigskip

\bigskip

\textbf{5. \ \ Role of knotted/linked null electromagnetic fields in the
theory }

\ \ \ \ \ \ \textbf{of nonrelativistic Schr\"{o}dinger equation}

\bigskip

Our readers at this point might ask a question: How the obtained results are
affected by currents and charges present in electromagnetic field? \ Recall,
Bohm \ [4,11] was concerned exactly with this issue when he compared \
quantum mechanical and optical formalisms. But currents are \ made of moving
charges! Therefore, the above question should be modified accordingly. In
[31, 32] we further developed results of Ranada [33,34] who skillfully using
electric-magnetic duality for \ source-free electromagnetic fields obtained
stable solutions which we reinterpreted as Dirac monopoles, electrons and
dyons (particles possessing simultaneously both the electric and magnetic
charges). Topologically all these objects are represented \ by the
interlinked (Hopf) rings made of closed electric (electron) or magnetic
(Dirac monopole) \ lines and, for dyons, by linked and closed \ two electric
and two magnetic lines. More complicated objects are also possible [32] but
dynamically they are less stable [35]. \ Many of these objects belong to the
so called null fields to be defined below but some are made of fields which
are not null [36]. The question remains: How the masses enter into this
purely topological \ picture? This issue was discussed in our works [31,35].
We argued that complements of knots and links make the ambient Minkowski
space-time curved in the same way as caused by masses in Einsteinian
gravity. \ Since masses should be positive, not all knots and links are
allowed to exist. This \ restriction excludes all hyperbolic knots and links
from consideration. Incidentally, mathematical methods described in [36] for
knot/link generation do not involve creation of hyperbolic knots/links. The
rationale for absence of hyperbolic knots and links in mathematical
formalism described in [36] was not explained. The same results as in [36]
were obtained in our work [32] where knots and links were generated
dynamically. This allowed us in the same paper to develop mathematical and
physical explanation of exclusion of hyperbolic knots and links from
consideration. Introduction of masses into general relativity as well as
charges into non Abelian Yang-Mills (NAYM) theory is associated with \ very
serious technical problems. These are described in [37], page 97, and
references therein. In the case on NAYM theories the problem \ of charges\
is by-passed by treating only source-free NAYM fields allowing existence of
monopoles, dyons, etc. [37]. In [31] we reobtained Ranada's results for
Abelian Yang-Mills fields using standard instanton formalism. At the same
time, methods for knot/link generation described in [36] do not involve uses
of instantons. Although it is commonly believed that instanton methods are
applicable only to the NAYM fields, many years ago Trautman obtained the
Dirac monopole solution with help of the Abelian YM fields [38].
Incidentally, the same result is obtained by using the NAYM fields. The
Abelian reduction method [31,37] permits to obtain results for the Dirac
monopole starting with the non Abelian t'Hooft-Polyakov monopole. This is
shown in detail in [39], page174. Since gravity can be rewritten in the
language of NAYM fields as it was shown in 1956 by Utiyama [40], it is only
natural that Robinson \ in 1961 was able to connect gravity with source-free
Maxwell's fields [36,41].

Now we would like to shed some light on the following issue: How presence \
or absence of masses is related to topics just described? Surprisingly, the
link between these topics can be established with help of \ the Huygens'
principle. This principle can be discussed purely mechanically with methods
of contact geometry and topology as it was done, for example, by Arnol'd
[42]. Reader's friendly basics of contact geometry and topology is provided
in [37]. The same principle could be discussed \ using theory of partial
differential equations discussed, for example, by Hilbert and Courant [17],
Luneburg [18], Hadamard [43] and later, by others. For the sake of space we
shall not go into details pertained to Huygens' principle in this Letter. We
only mention that \ equations (31a,b) do obey the Huygens' principle so that
all equations which can be reduced to (31 a,b) \ do obey this principle.
According to Hadamard, only 3 operations are allowed for conversion of a
given PDE equation to the "trivial" equation(s) (31a,b). Specifically, the 
\textsl{Huygens' equivalence principle} \ can be formulated as follows.

Let L[$\phi $] be the Huygens-equivalent operator (that is operator which is
equivalent to (31a,b)) and let \~{L}[$\phi $] be another operator. Then,
they are Huygens-equivalent if :

a) \~{L}[$\phi $] can be obtained from L[$\phi $] by the nonsingular
transformations of independent

\ \ \ \ variables.

b) \~{L}[$\phi $]=$\lambda ^{-1}$L[$\lambda \phi $] for some positive,
smooth function $\lambda $ of independent variables.

c) \~{L}[$\phi $]=$\rho $L[$\phi $] for some positive smooth function $\rho $
of independent variables.

With these rules in our hands, let us consider now an invertible sequence of
transformations from the d'Alembert equation (31) to the massive K-G
equation. Let us use the transformation \~{L}[$\phi $]=$\lambda ^{-1}$L[$%
\lambda \phi $] with $\lambda =e^{\alpha t},$ with $\alpha $ some constant.
Then, we obtain:%
\begin{equation*}
e^{-\alpha t}\{[\frac{\partial ^{2}}{\partial t^{2}}e^{\alpha t}\phi
]-[\nabla ^{2}e^{\alpha t}\phi ]\}=\frac{\partial ^{2}}{\partial t^{2}}\phi
+2a\frac{\partial }{\partial t}\phi -\nabla ^{2}\phi +a^{2}\phi =0.
\end{equation*}%
Next, let $\lambda _{1}=e^{ibx}$ and $\lambda _{2}=e^{-icx}.$ Upon
substitution of these factors into previous equation, while using the rule
b), we obtain after some calculation 
\begin{equation*}
\frac{\partial ^{2}}{\partial t^{2}}\phi +2a\frac{\partial }{\partial t}\phi
-\nabla ^{2}\phi +2ib\frac{\partial }{\partial x}\phi -2ic\frac{\partial }{%
\partial x}\phi +(a^{2}-b^{2}-c^{2})\phi =0.
\end{equation*}%
If now $b=c$ and $a^{2}=2b^{2},$ we obtain the telegrapher equation%
\begin{equation*}
\frac{\partial ^{2}}{\partial t^{2}}\phi +2a\frac{\partial }{\partial t}\phi
-\nabla ^{2}\phi =0.
\end{equation*}%
The K-G equation is obtained now upon substitution $\phi =e^{mt}\psi $ into
telegrapher's equation with subsequent replacement of $m$ by $im$. Thus the
K-G and \ d'Alembert equations are Huygens' -equivalent. If this is so, it
would be possible to use the twistor formalism [36] for the K-G equation.
The question arises: Is the stationary Schr\"{o}dinger equation
Huygens-equivalent to the d'Alembert equation? \ 

In 1935 Fock initiated study of this problem (having different goals in mind
though). \ While studying \ the (accidental) degeneracy of the stationary
Schr\"{o}dinger equation for the hydrogen atom he converted this equation
into the integral equation looking very similar to the Poisson integral in
the theory of \ functions of one complex variable. Recall that harmonic
functions (that is functions obeying the Laplace equation) inside the circle
(and, thus, of any domain which can be conformally mapped into circle) are
represented via Poisson integral. Fock initiated study of what is known now
as dynamical symmetry groups. These groups are allowing to solve quantum
mechanical problems group-theoretically thus by-passing uses of the Schr\"{o}%
dinger equation. This \ direction of research has grown into a large field
[44] nowadays. Fock's work is presented in Vol.1, pages 400-410 of [44].
Group-theoretic analysis had lead Fock to conclusions that the bound states
of hydrogen atom should be studied in 4-dimensional Euclidean space while
scattering states should be studied in 3+1 dimensional hyperbolic
(Lobachevski) \ space. The Euclidean 4-dimensional version of the Schr\"{o}%
dinger equation (for the bound states) was reduced by Fock to study of
solutions of the 4-dimensional Laplacian. The description of the scattering
states was only outlined by Fock. He suggested that this case would require
study of solutions of the d'Alembertian. This suggestion was waiting for its
solution \ for 31 years. In 1966 it was finally solved by Itzykson and
Bander [45]. At the level of classical mechanics all finite dimensional
exactly integrable systems are equivalent to each other since their
trajectories in phase space are Liouville tori [42]. Thus, Fock results for
the hydrogen atom are canonically extendable to all two--body exactly
solvable nonrelativistic mechanical problems. Although Itzykson and Bander
[45] obtained the d'Alembertian for scattering states thus making the
stationary Schr\"{o}dinger equation Huygens-equivalent to the d'Alembertian
equation, \ the results of [45] happened to be very cumbersome. The
transparency of results for both bound and scattering states of hydrogen
atom was achieved \ only in 2008 in the paper by Frenkel and Libine [46].
Using quaternions-analogs of complex numbers in 4 dimensions- these authors
extended the Poisson formula for harmonic functions from 2 to 4 dimensions.
This allowed them to extend effortlessly the obtained results from Euclidean
4 dimensional to Minkowski 3+1 spaces.

With these results in our hands now we are in a position to explain quantum
mechanical significance of the null fields. These fields were known \ since
\ 1787, when Chladni, a German physicist, \ studied nodal lines of a
vibrating metal plate by stroking this plate covered by sand with a violin
bow [47]. After studying a variety of nodal patterns systematically, he
wrote a book in 1802 where all these patterns were systematized. Remarkably,
the 1802 edition of Chladni book was translated into English and published
in 2015 [48]. Nowadays, Chladni patterns can be seen on You Tube [49].

\ With help of results of previous sections now we are in the position to
conclude this Letter \ with explanation of the connection between Chladni \
patterns, null fields, knots/links and quantum mechanics. To set up the
notations, we begin with the description of two dimensional Chaladni
problems following [50]. \ It is convenient to think about a given
Riemannian surface $\Sigma $ with metric $g_{\Sigma }$ as a vibrating
membrane with $u(\mathbf{x},t)$, ($\mathbf{x}\in \Sigma )$ being a
displacement\ at time $t$ of the membrane from its original position. The
function $u$ is a solution to the wave equation 
\begin{equation}
\frac{\partial ^{2}}{\partial t^{2}}u=\nabla _{\Sigma }^{2}u.  \tag{43}
\end{equation}%
By representing solution in the form \ $u(\mathbf{x},t)=v(t)w(\mathbf{x})$,
the above equation splits into two equations, e.g. 
\begin{equation}
\frac{\partial ^{2}}{\partial t^{2}}v=\lambda v  \tag{44a}
\end{equation}%
and%
\begin{equation}
\nabla _{\Sigma }^{2}w=\lambda w.  \tag{44b}
\end{equation}%
The (null) \textsl{zero set}, $\Xi (w):=\{\mathbf{x\in \Sigma :}$ $w\mathbf{%
(x)=}0\},$ is called \textsl{nodal} (Chladni) set. The definitions \ just
described need to be supplemented with the boundary (e.g. Dirichlet or
Neumann) conditions so that the above nodal problem is known as \textsl{%
fixed membrane problem}. In the case when the membrane is a closed surface
the above problem is known as\textsl{\ free membrane problem. }The first%
\textsl{\ \ }detailed calculation of Chladni patters was\textsl{\ }\ done by
Poisson in \ 1829. The chronology of \ subsequent developments along with
detailed numerous examples \ can be found in encyclopedic book [51] by Lord
Rayleigh in ch.-rs 9 and 10. The same results were recently reproduced in
the book [52]. At much more advanced level Chladni patters were studied by
Cheng\textsl{\ }[53] who proved that for an arbitrary smooth Riemannian
surface ($\Sigma ,g_{\Sigma })$ the nodal set is a collection of immersed
closed curves. These curves are easily seen in [49] at sufficiently high
frequencies, provided the periodic boundary conditions on the plate are
imposed. Cheng's result is remarkable since later on analogous results were
obtained in 3 dimensions in [54]. This time, since the nodal curves are
closed, \ they can generate knots and links, of any complexity. The results
of [54] \ can be reobtained with help of results obtained in this Letter
superimposed with [31,32,37]. Following [55] we write%
\begin{equation}
\mathbf{F}(\mathbf{r},t)=\mathbf{F}_{+}+\mathbf{F}_{-}.\text{ \ }\mathbf{F}%
_{\pm }(\mathbf{r},t)=\dint\limits_{0}^{\infty }d\omega e^{\mp i\omega t}%
\mathbf{F}_{\pm \omega }(\mathbf{r}).  \tag{45}
\end{equation}%
In such notations (7),(8) can be rewritten as

\bigskip 
\begin{eqnarray}
\mathbf{\nabla }\cdot \mathbf{F}_{\pm } &=&0,\text{ }  \TCItag{46a} \\
\nabla \times \mathbf{F}_{\omega } &=&k\mathbf{F}_{\omega }  \TCItag{46b}
\end{eqnarray}%
with $k=\omega /c.$ In plasma physics (46b) is known as "force-free
equation" while in \ hydrodynamics it is known as "Beltrami" equation. It
was discussed in detail in our book [37] in the context of methods of
contact geometry/topology while in [31] and [32] these \ methods were used
for generation of all kinds of nonhyperbolic knots and links, including
those of Ranada-type, that is Hopf-like links. Applying operator $\mathbf{%
\nabla }\times $ \ to (46b) while taking into account (46a) results in 
\begin{equation}
\nabla ^{2}\mathbf{F}_{\omega }+k^{2}\mathbf{F}_{\omega }=0  \tag{47}
\end{equation}%
to be compared with (44).\ This\ (vector) version of the Helmholtz equation
is known as Chandrasekhar-Kendall (CK) equation [37]. These authors noticed
that every solution of (46b) is solution of (47) but the converse is not
true. This happens to be of fundamental importance for our tasks. In [37],
page 30, we stated that solution \ of (46b) is a composition of fields of 3
types: a) solenoidal (46a), b) lamellar $\mathbf{F}_{\omega }\cdot \left(
\nabla \times \mathbf{F}_{-\omega }\right) =0,$c) Beltrami $\mathbf{F}%
_{\omega }\times \left( \nabla \times \mathbf{F}_{-\omega }\right) =0.$ The
null fields used in creation of "linked and closed beams of light" [36] are
lamellar. In terms of notations set up in [55] they are defined as 
\begin{equation}
\mathbf{F}_{\omega }\cdot \mathbf{F}_{-\omega }=0.  \tag{48a}
\end{equation}%
As explained in [37], the same classification of fields exist in physics of
liquid crystals. In [37], pages 32-34, it was demonstrated that the
Faddeev-Skyrme (F-S) knot/link generating model is of the same liquid
crystalline origin. It also can be looked upon as originating from the
Abelian reduction of NAYM fields. Therefore, all results of [36] \ as well
as of [31,32] are compatible with those originating from the F-S model [56].
Now we can study lamellar (null) solutions of the C-K equation. Being
logically guided by [54,55] our results differ in details from results of
these works.4sc This permits us to inject new physics absent in these
references. We begin with (48a). It can be rewritten as 
\begin{equation}
\left\vert \mathbf{E}_{\omega }\right\vert ^{2}-\left\vert \mathbf{H}%
_{\omega }\right\vert ^{2}+2(\mathbf{E}_{\omega }\cdot \mathbf{\bar{H}}%
_{\omega }+\mathbf{\bar{E}}_{\omega }\cdot \mathbf{H}_{\omega })=0. 
\tag{48b}
\end{equation}%
This \ equation is surely satisfied if $\left\vert \mathbf{E}_{\omega
}\right\vert ^{2}=\left\vert \mathbf{H}_{\omega }\right\vert ^{2}$ and $%
\mathbf{E}_{\omega }\cdot \mathbf{\bar{H}}_{\omega }+\mathbf{\bar{E}}%
_{\omega }\cdot \mathbf{H}_{\omega }=0.$Next, without loss of generality and
following [55] it is permissible to assume that $\mathbf{E}_{\omega }=%
\mathbf{\bar{E}}_{\omega }$ and $\mathbf{H}_{\omega }=\mathbf{\bar{H}}%
_{\omega }.$ This then leads to $\mathbf{E}_{\omega }\cdot \mathbf{H}%
_{\omega }=0.$ Should we avoid use of the time Fourier transforms given \ in
(45), just presented two (null) equations (that is $\left\vert \mathbf{E}%
_{\omega }\right\vert ^{2}=\left\vert \mathbf{H}_{\omega }\right\vert ^{2}$
and $\mathbf{E}_{\omega }\cdot \mathbf{H}_{\omega }=0)$ are sufficient for
generation of all kinds of torus knots and Hopf links, including those
discovered by Ranada, evolving in time [36, 57,58]. These results do not let
us to make a connection with Chladni patterns though and with the physics
associated \ with these patterns. To correct this, using\ [59] we consider
the following remarkable identity%
\begin{equation}
(\nabla ^{2}+k^{2}\mathbf{)(r\cdot v)}=2\nabla \mathbf{v}+\mathbf{r\cdot }%
(\nabla ^{2}+k^{2}\mathbf{)v.}  \tag{49}
\end{equation}%
Here $\mathbf{v}$ is either $\mathbf{E}_{\omega }$ or $\mathbf{H}_{\omega }$%
. The scalar $\mathbf{(r\cdot F}_{\omega }\mathbf{)}$ is convenient to
rewrite in the notations of [54], i.e. $\mathbf{(r\cdot F}_{\omega }\mathbf{%
)=}u=u_{1}+iu_{2}.$ Evidently, in view of (16b) scalar $u$ can be identified
with $V$(\textbf{x},t) defined in (13) so that $\mathbf{E}_{\omega }$ and $%
\mathbf{H}_{\omega }$ can be recovered if needed. Fortunately, this will not
be necessary in the rest of this Letter. By combining $(47)$ and $(49)$ we
obtain Chladni-type equations 
\begin{equation}
(\nabla ^{2}+k^{2}\mathbf{)}u_{_{k}1}=0\text{ and }(\nabla ^{2}+k^{2}\mathbf{%
)}u_{_{k}2}=0.  \tag{50}
\end{equation}%
These equations were introduced and discussed in [54] using purely
mathematical arguments. Since here the same equations were reobtained with
help of physical arguments, this allows us to extend results of [54]. First,
we notice that both equations have \textsl{the same eigenvalue} $\lambda
_{1}=\lambda _{2}=k^{2}$. Having the same eigenvalue (of multiplicity 2) the
wavefunctions $u_{_{k}1}$ and $u_{_{k}2}$ are not the same though as
demonstrated in [54] and below. Although the presence of i-factor is
essential, the difference goes beyond \ this fact. This difference should
not be confused with the degeneracy concept in quantum mechanics. Presence
of eigenvalues having multiplicities is responsible for effects of
entanglements. This is explained in detail our book [37], pages 386-395. Now
we are in the position to demonstrate that: a) quantum mechanical
entanglement is equivalent with the entanglement in the topological
knot-theoretic sense; b) presence of eigenvalues with multiplicity makes Schr%
\"{o}dinger's and Heisenberg's interpretation of quantum mechanics not
equivalent. The last statement follows immediately from the detailed
Heisenberg-style calculations presented in [60]. Therefore, we are only left
with explanation of a).

Following [31] consider force-free equation (46b), where temporarily we
replace $\mathbf{F}_{\omega }$ by $\mathbf{v}$ as in (49). Then by applying
to both sides of (46b) \ the operator div and \ by assuming that $%
k=const=\kappa (x,y,z)$ we obtain div($\kappa \mathbf{v)}=\mathbf{v}\cdot
\nabla \kappa =0.$ Let $\mathbf{r}(t)=\{x(t),y(t),z(t)\}$ be some trajectory
on the surface $const=\kappa (x,y,z).$ In such a case $\frac{d}{dt}\kappa
\{x(t),y(t),z(t)\}=v_{x}\kappa _{x}+v_{y}\kappa _{y}+v_{z}\kappa _{z}=%
\mathbf{v}\cdot \nabla \kappa =0.$ This means that the "velocity" $\mathbf{v}
$ is always tangential to the surface $const=\kappa (x,y,z).$ Since the
vector field $\mathbf{v}$ is being assumed nowhere vanishing, the surface $%
const=\kappa (x,y,z)$ can only be as torus T$^{2}.$ The field lines of 
\textbf{v} on T$^{2}$ should be closed if $const$ is rational number. Thus
we just demonstrated that the force-free equation (46b) supplies us with the
condition of existence of all possible torus knots for rational $\kappa
^{\prime }s$ . Starting with work by Ranada such torus knots were explicitly
designed both in [31, 32] and [36]. Now it remains to demonstrate that
Chladni-type equations (50) \ lead to 3-dimensional Chladni patterns
associated with these equations. Since this was done already in [54], our
task is only to supply some physics to the results of [54]. Since the
surface $const=\kappa (x,y,z)$ is T$^{2}$, while the solid torus is defined
by \^{T}$^{2}=$ $D^{2}$ $\times S^{1},$ with $D^{2}$ being a disc, it is
convenient to introduce a cylindrical system of coordinates and to consider
the Neumann-type problem\ for the Helmholtz equations (50) written in
cylindrical coordinates. This task is facilitated by the accumulated
knowledge about circular waveguides in electrodynamics.\ The solutions $%
u_{_{k}1}=J_{1}(\rho \sqrt{\tilde{\lambda}})\cos \varphi $ and \ $u_{_{k}2}$
=$J_{1}(\rho \sqrt{\tilde{\lambda}})\sin \varphi $\ discussed in [54]%
\footnote{%
Here $J_{1}(x)$ is the standard Bessel function and $\tilde{\lambda}$
adjusted (with account of cylindrical symmetry ) egenvalue .} are \ just TE
and TM -type solutions \ known for circular waveguides [61].\ \ For small $%
\rho ^{\prime }s$ \ they are represented by \ \ $u_{_{k}1}\simeq \frac{\sqrt{%
\tilde{\lambda}}}{2}\rho \cos \varphi $ and \ \ $u_{_{k}2}\simeq \frac{\sqrt{%
\tilde{\lambda}}}{2}\rho \sin \varphi $\ . \ Both functions become zero for $%
\rho =0.$ Thus, the individual nodal Chladni sets are respectively given by $%
u_{_{k}1}^{-1}(0)$\ \ and $u_{_{k}2}^{-1}(0),$\ while the Chladni
centerline\ $S^{1\text{ }}$ for the solid torus\ is determined by the
transversality condition: $u(0)=u_{_{k}1}^{-1}(0)\cap $\ $u_{_{k}2}^{-1}(0).$
It is indeed the transversality condition since we can plot both $\rho \cos
\varphi $ and $\rho \sin \varphi $ on\ the complex plane \textbf{C} \ so
that they are transversal to each other [62].The transversality condition
needed for the validity of Thom's isotopy theorem [54,62] (assuring that the
obtained results are stable with respect to possible perturbations). These
perturbations will occur because \ by definition every knot is an embedding
of \ $S^{1}$ into $S^{3}$ $=\mathbf{R}^{3}\cup \{\infty \}.$ The validity of
Thom's theorem is required \ to assure that the embedding of solid tori \^{T}%
$^{2}$ into $S^{3}$ could be done in such a way that it shall produce knots
of any complexity as long as they are non wild. \ Although [54] claims that
all knots could be obtainable this way, physically, this is unrealistic
since complex knots are dynamically generated from \ the simpler ones [35]
and such process precludes the formation of hyperbolic-type knots [32].

\bigskip

\textbf{References}

\bigskip

[1] \ R.Feynman and R.Leighton, The Feynman Lectures on Physics, Vol.3, 

\ \ \ \ \ Basic Books, New York, 2011.

[2] \ M.Arndt, O.Nairz,J.Vos-Andreae, C.Keller, G.Zouw and A.Zellinger, 

\ \ \ \ \ \ Nature, 401 (1999) 680.

[3] \ S.Eibenberger, S. Gerlich, M.Arndt, M. Mayor and J. T\"{u}xen, \ Phys. 

\ \ \ \ \ \ Chem. Chem. Phys. 15 (2013) 14696.

[4] \ D.Bohm, Quantum Theory, Dover Publications Inc., New York, 1989.

[5] \ A. Sanz, and S. Miret-Art\'{e}s, A Trajectory Description of Quantum 

\ \ \ \ \ \ Processes. I. Fundamentals, Springer-Verlag,Berlin, 2012.

[6] \ U.Leonhardt, Measuring the Quantum State of Light, Cambridge U. 

\ \ \ \ \ \ Press, Cambridge, UK, 1997.

[7] \ M. Fox, Quantum Optics: An Introduction, Oxford U. Press, 

\ \ \ \ \ \ Oxford,UK, 2006.

[8] \ V. Berestetskii, E. Lifshitz and L.Pitaevskii, Relativistic Quantum 

\ \ \ \ \ \ Theory, Pergamon Press, Oxford, UK, 1971.

[9] \ E.Collett, Am.J.Phys. 39, (1971) 1483.

[10] B.Kanseri, N.Bisht and H.Kandpal, Am. J.Phys.76 (2008) 39.

[11] D.Bohm and B.Hiley, Undivided Universe, Rutlege Publ.Co., London, 1993.

[12] D.Kobe, Found.Phys. 29 (1999)1203.

[13] M.Raymer and B. Smith, SPIE Conference on Optics and Photonics,

\ \ \ \ \ \  Conference number 5866, (2005).

[14] M.Raymer and B. Smith, New J.Phys. 9 (2007) 414.

[15] H.Green and E.Wolf, Proc.Phys.Soc. A 66 (1953) 1129.

[16] E.Wolf, Proc.Phys.Soc. A 74 (1959) 269.

[17] D. Hilbert and R.Courant, Methods of Mathematical Physics, Vol.2,

\ \ \ \ \ \  Interscience Publishers, New York, 1962.

[18] R.Luneburg, Mathematical Theory of Optics, U.of California Press,

\ \ \ \ \ \  Berkeley, CA, 1966.

[19] V. Maslov and M.Fedoriuk, Semiclassical Approximation in Quantum

\ \ \ \ \ \ \  Mechanics, Reidel Publ. Co., Boston, MA, 1981.

[20] H.R\"{o}mer, Theoretical Optics, Wiley-VCH, Hoboken, NJ, 2005.

[21] A. Akhiezer and V. Berestetsky, Quantum Electrodynamics, Interscience

\ \ \ \ \ \  Publishers, New York, 1965.

[22] N.Bogoliubov and D.Shirkov, Introduction to the Theory of Quantized
Fields, 

\ \ \ \ \ \ \ John Wiley\&Sons, New York,1976.

[23] Harish-Chandra, Proc.Roy.Soc.London, A186 (1946) 502.

[24] P.Ghose, A.Majumdar, S.Guha and J.Sau, Phys.Lett.A 290 (2001) 205.

[25] B.Shabat, Introduction to Complex Analysis.Part II, American
Mathematical 

\ \ \ \ \ \ \ Society, Providence, RI, 1992.

[26] P. Roman, Proc.Phys.Soc. A 74 (1959) 281.

[27] W.Greiner, Relativistic Quantum Mechanics.Wave Equations, 

\ \ \ \ \ \ \ Springer-Verlag, Berlin, 2000.

[28] Z.Tokuoka, Progr.Theor.Phys. 10 (1953) 137.

[29] E. Rojas, B.El-Bennich, J. de Melo and M. Paracha, Few-Body Syst. 56
(2015) 639.

[30] E.Schr\"{o}dinger, Collected Papers on Wave Mechanics,Chelsea Publ.Co.,

\ \ \ \ \ \ \ New York, 1978.

[31] A.Kholodenko, Analysis \& Math. Phys, 6 (2016)163.

[32] A.Kholodenko, Ann.Phys. 371 (2016) 77.

[33] A.Ranada, Lett. Math. Phys. 18 (1989) 97.

[34] A.Ranada, J.Phys. A 25 (1992) 1621.

[35] A.Kholodenko, Int. J.Mod.Phys.A 30 (2015) 1550189.

[36] A.Arrayas, D.Bouwmeester and J.Trueba, Phys.Rep. 667 (2017) 1.

[37] A.Kholodenko, Applications of Contact Geometry and Topology in Physics, 

\ \ \ \ \ \ \ World Scientific, Singapore, 2013.

[38] A.Trautman, Int.J.Theor.Phys. 16 (1977) 561.

[39] M.G\"{o}ckeler and T.Shuker, Differential Geompetry, Gauge Theories and
Gravity, 

\ \ \ \ \ \ \ Cambr.U.Press, Cambridge, UK, 1987.

[40] R.Utiyama, Phys.Rev. 101 (1956) 1597.

[41] I.Robinson, J.Math.Phys.2 (1961) 290.

[42] V. Arnol'd, Mathematical Methods of Classical Mechanics,
Springer-Verlag, 

\ \ \ \ \ \ \ Berlin, 1989.

[43] P.G\"{u}nter, Huygens Principle and Hyperbolic Equations, Academic
Press Inc.,

\ \ \ \ \ \ \ Boston, MA, 1988.

[44] A.Bohm, Y.Ne'eman and A.Barut, Dynamical Groups and Spectrum Generating

\ \ \ \ \ \  Algebras, Vol's 1 \& 2, World Scientific, Sigapore, 1988.

[45] C.Itzykson and M.Bander, Group theory of the hydrogen atom, I \&II,

\ \ \ \ \ \  Rev.Mod.Phys. 38 (1966), 330 and 346.

[46] I.Frenkel and M.Libine, Adv.Math. 218 (2008) 1806.

[47] D. Jacobson, N.Nadirashvili and J.Toth, Russian Math.Surveys 56 (2001)
1085.

[48] E.Chaladni, Tretease on Acoustic, Springer-Verlag, Berlin, 2015.

[49] https://www.youtube.com/watch?v=3idd3GEMoWk

[50] R. Komendarczyk, AMS Transactions,358 (2005) 2399.

[51] J.Rayleigh, Theory of Sound, Vol.1, Macmilland and Co. Ltd, London,
1896.

[52] T.Rossing and N.Fletcher, Principles of Vibrations and Soundm
Springer-Verlag, 

\ \ \ \ \ \ \ Berlin, 2004.

[53] S-Y. Cheng, Comm.Math.Helvetici 51 (1976) 43.

\bigskip \lbrack 54] A.Enciso, \ D.Hartley and D.Peralta-Salas,
J.Funct.Analysis 271 (2016) 182.

[55] G.Kaiser, J.Opt. A 6 (2004) S243.

[56] http://hopfion.com/faddeev.html

[57] I.Besieris and A. Shaarawi, Optics Lett. 34 (2009) \ 3887.

[58] J.Waite, The Hopf Fibration and Encoding Torus Knots into Light Fields,

\ \ \ \ \ \ \ \ Ms of Sci. thesis, U.of Nevada, 2016.

[59] C. Bouwkamp and H. Casimir, Physica 20 (1954) 539.

[60] A.Kholodenko, Int.Math.Forum 4 (2009) 509.

[61] http://nptel.ac.in/courses/112105165/lec14.pdf.

[62] A.Majthay, Foundations of Catastrophe Theory, \ Pitman Publishers,
Boston, 1985.

\bigskip

\bigskip

\bigskip

\bigskip

\bigskip

\bigskip

\bigskip

\bigskip

\bigskip

\end{document}